\pdfoutput=1

\documentclass[
    ,final            
  ]
  {aipproc}

\layoutstyle{8x11double}

\def\lsim{\raise0.3ex\hbox{$<$\kern-0.75em\raise-1.1ex\hbox{$\sim$}}}
\def\gsim{\raise0.3ex\hbox{$>$\kern-0.75em\raise-1.1ex\hbox{$\sim$}}}
\begin{document}

\title{Lattice results on the phase structure and equation of state in QCD at finite temperature}


\classification{11.15.Ha, 12.38.Gc, 12.38.Mh, 21.65.Qr, 25.75.Nq}
\keywords      {lattice QCD, finite temperature, QCD transition temperature, chiral scaling, equation of state}

\author{Kazuyuki Kanaya}{
  address={Graduate School of Pure and Applied Sciences, University of Tsukuba, 
        Tsukuba 305-8571, Japan}
}

\begin{abstract}
I review recent developments in the studies of the phase structure and equation of state in finite temperature QCD on the lattice. 
\end{abstract}

\maketitle



The rapid progress of heavy ion collision experiments urge us towards quantitatively reliable determination of these properties. 
Lattice QCD is a powerful basis to investigate thermodynamic properties of the quark matter around the deconfinement transition temperature directly from the first principles of QCD.
There were quite a few big advances in finite temperature QCD on the lattice made in the last years. 

Because the $s$ quark mass is comparable with the QCD scale $\Lambda_{\rm QCD}$ and thus the transition temperature,
it is important to incorporate the $s$ quark dynamically for quantitative predictions. 
Large-scale simulations in 2+1 flavor QCD with various improved staggered quark actions have now been started to produce results for various thermodynamic quantities which are extrapolated to the continuum limit at around physical quark masses.
At the same time, the theoretical uneasiness with staggered quarks motivated several groups to accelerate studies with Wilson-type quarks and with lattice chiral quarks. 
I review these recent developments in finite temperature lattice QCD, concentrating on the topics of the phase structure and the equation of state.
See \cite{KanayaLat10} for more discussions.


\section{Transition temperature}

The transition temperature $T_c$ is one of the most important quantities for experimental investigations of QGP.
Estimation of $T_c$ in 2+1 flavor QCD has been made based on large-scale simulations using various improved staggered quarks. 
However, there has been a sizable discrepancy in $T_c$ among groups for more than five years.

\paragraph{The discrepancy}

In 2005, the MILC Collaboration obtained $T_c=169(12)(4)$ MeV in the chiral and continuum limits from a study of the chiral susceptibility using asqtad quarks \cite{asqtad} on $N_t=4$--8 lattices \cite{MilcPRD71}, where the scale was set by $r_1$ and the O(4) critical exponents were adopted in the chiral extrapolation.
In 2006, based on a simulation using stout quarks \cite{stout} on $N_t=6$--10 lattices, the Wuppertal-Budapest Collaboration published 
$T_c = 151(3)(3)$ MeV from the susceptibility of the chiral condensate
and 
$175(2)(4)$ MeV from the $s$ quark number susceptibility \cite{WB_Tc1}.
These are the results extrapolated to the physical point and to the continuum limit, 
and the scale was set by $f_K$.
In the same year, the BNL-Bielefeld Collaboration published $T_c = 192(7)(4)$ MeV at the physical point in the continuum limit,
based on a study using p4 quarks \cite{p4}  on $N_t=4$ and 6 lattices \cite{BNLB_Tc1}.
This $T_c$ is an average of the values from the Polyakov-loop susceptibility and the chiral susceptibility, and the difference between them is included in the systematic error.
Later, the HotQCD Collaboration (a merger of the BNL-Bielefeld and MILC Collaborations) compared p4 and asqtad actions and found that the two staggered quark actions lead to roughly consistent $T_c$ on finite lattices, while it shifts towards smaller values when the lattice spacing is decreased and when the light quark mass is decreased \cite{BNLB_EOSa,BNLB_EOSt8, BNLB_EOSp}.

Because the transition is considered to be an analytic crossover around the physical point (see e.g.\ \cite{WB_Nature}), the value of $T_c$ may depend on the choice of observable to define it.
$T_c$ should depend on the details of the analyses, such as the scale convention, too. 
However, it was found that the discrepancy exists even when we adopt the same observable and the same scale convention.

\paragraph{Resolution}

\begin{figure}[tbh]
  \includegraphics[width=0.24\textwidth]{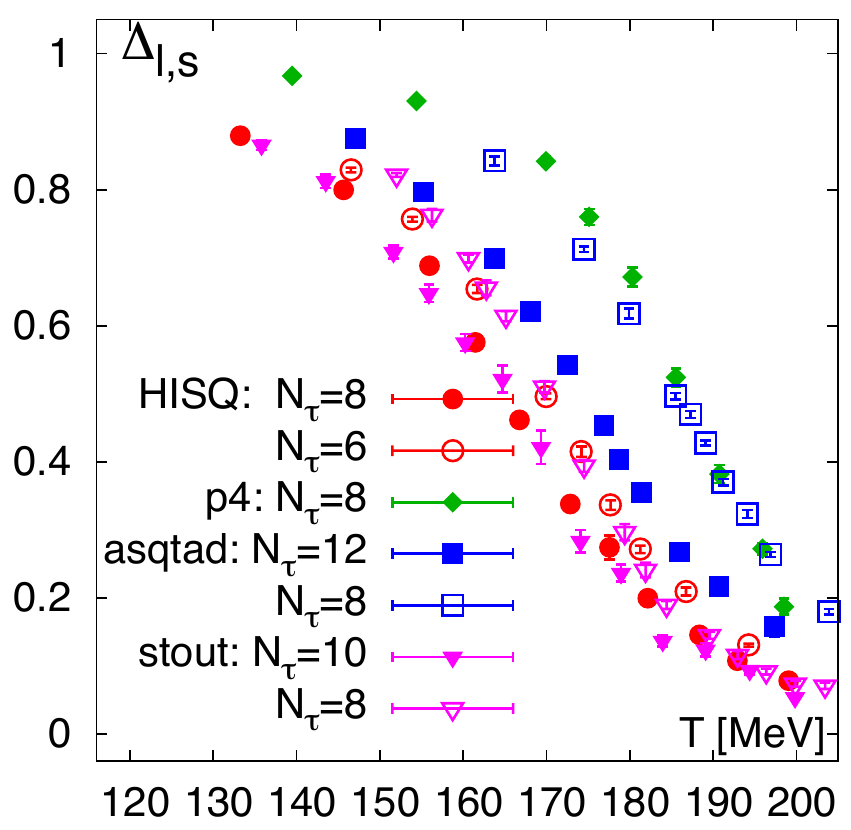}
\hspace{5mm}
  \includegraphics[width=0.275\textwidth]{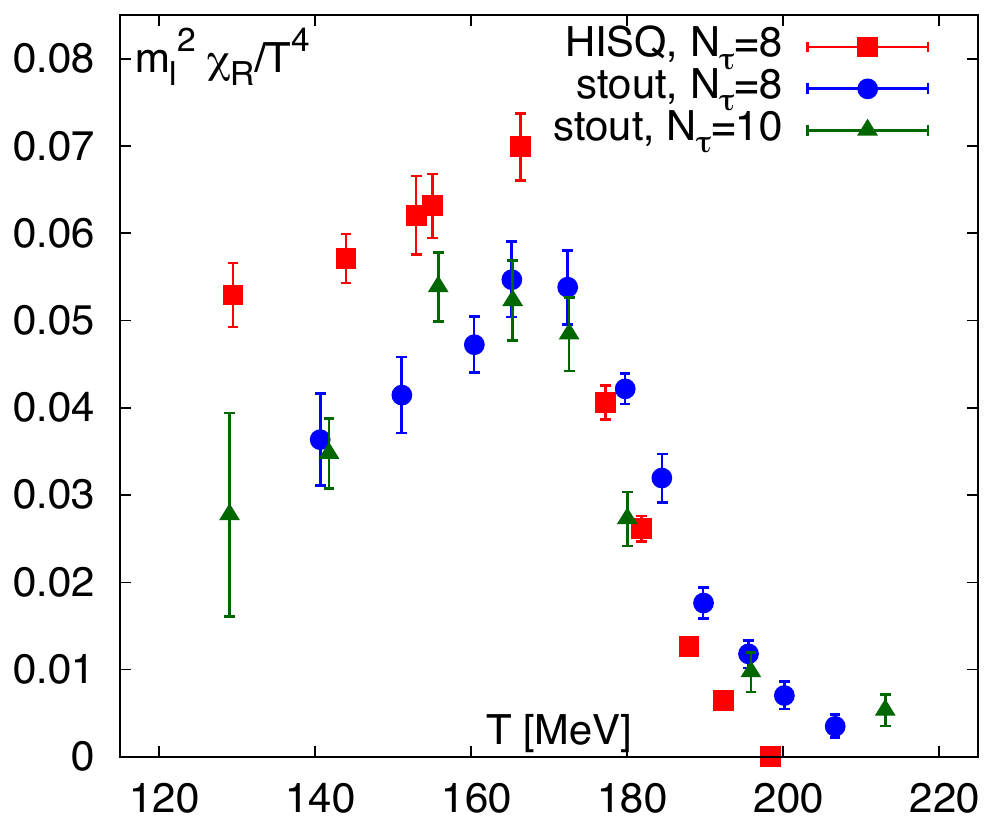}
\hspace{4mm}
  \includegraphics[width=0.275\textwidth]{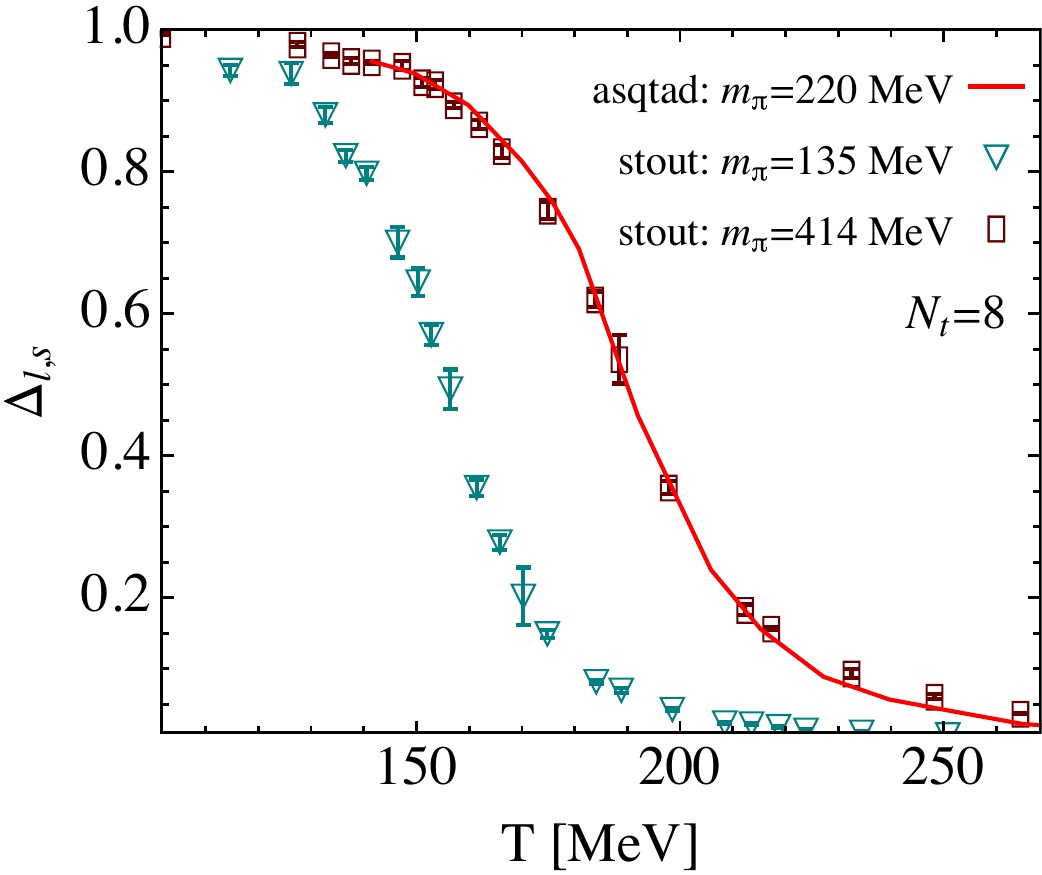}
\caption{
Chiral observables in 2+1 flavor QCD with improved staggered quarks.
{\bf (Left)}
Subtracted chiral condensate (difference of scaled light quark and $s$ quark condensates to remove a divergent renormalization factor) from the HISQ, p4, asqtad and stout actions \cite{BazavovSoeldnerLat10}.
{\bf (Center)}
Comparison of HISQ and asqtad results for the disconnected part of the chiral susceptibility \cite{BazavovSoeldnerLat10}.
In these panels, the HISQ, p4 and asqtad data are obtained at the bare quark mass ratio $m_{ud}/m_s = 0.05$ with $m_s$ around the physical value, using a scale fixed by $r_0$ .
The stout data, which is obtained at the physical point using a $f_K$ scale \cite{WB_Tc3}, is shifted in $T$ to correct the difference in the scale setting.
Note that this procedure causes a slight deviation from the physical point at finite lattice spacings.
{\bf (Right)} %
Subtracted chiral condensate with stout and asqtad quarks at $N_t=8$ \cite{WB_Tc3}. 
The open rectangles are results of stout quark action at $m_\pi^{\rm pNG}=414$ MeV \cite{WB_Tc3},  and the red curve is the result of asqtad quark action at $m_\pi^{\rm pNG}=220$ MeV \cite{BNLB_EOSt8}.
The stout quark mass for open rectangle is adjusted to reproduce $m_\pi^{\rm RMS}=587$ MeV of the asqtad data at $T\approx 135$ MeV on an $N_t=8$ lattice.
}
\label{Fig:Tc1}
\end{figure}

The Wuppertal-Budapest Collaboration extended the study to finer lattices ($N_t$ up to 16) and concluded that their values remain essentially unchanged \cite{WB_Tc2,WB_Tc3}.
Their final values are 
$T_c = 147(2)(3)$ MeV from the susceptibility of the chiral condensate 
and $T_c = 165(2)(4)$ MeV from the $s$ quark number susceptibility \cite{WB_Tc3}.
At the Lattice 2010 conference, the HotQCD Collaboration presented a preliminary result $T_c=164(6)$ MeV at the physical point in the continuum limit, 
based on a study of the disconnected chiral susceptibility using asqtad quarks on lattices up to $N_t=12$ \cite{BazavovSoeldnerLat10}.
In parallel, they presented first results of their new project \cite{BazavovSoeldnerLat10,BazavovPetrecky} adopting the ``highly improved staggered quark (HISQ)'' action \cite{HISQ}.
Comparison of these actions is shown in left and center panels of Fig.~\ref{Fig:Tc1}.
Consulting new results, 
we find that the discrepancy in $T_c$ is now mostly removed: 
All the actions converge towards $T_c \approx145$--165 MeV.

\paragraph{Staggered quarks and the taste violation}

What was the origin of the discrepancy?
At this point, I have to comment on several caveats with staggered-type lattice quarks.
A naive lattice discretization of the Dirac action leads to the species doubling problem.
To avoid this, various lattice fermions have been proposed, including the Wilson-type, the staggered-type and lattice chiral fermions.
Among them, most quantitative outputs of finite temperature QCD have been produced with staggered-type lattice quarks because of their relatively cheap simulation costs. 

On the other hand, the staggered-type quarks lead to four copies (``tastes'') of fermions for each flavor in the continuum limit.
To remove unwanted three tastes, we usually adopt the ``fourth root procedure'', i.e.\ we replace the quark determinant by its fourth root in the path-integration procedures.
Since this makes the theory non-local and non-unitary, many theoretical issues emerge \cite{rStag1}, as intensively discussed in the Confinement 8 workshop.
In particular, the universality arguments become fragile due to non-local interactions, and thus, e.g.\ we are not fully sure if the rooted staggered quarks have the correct continuum limit,
though, so far, no apparent obstacles about the validity of the rooting procedure seem to exist, provided that the continuum extrapolation is done prior to the chiral extrapolation \cite{rStag1,rStag4}.

Even when we assume that the staggered quarks do have the correct continuum limit of QCD with desired number of flavors,
a couple of issues remain on finite lattices.
One of them is the problem of taste violation.
The symmetry among tastes in the continuum limit is explicitly broken at finite lattice spacings.
This introduces a systematic uncertainty in the identification of flavors, 
and we observe different masses for a hadron depending on the combination of tastes.
For the pion, one combination of tastes, the pseudo Nambu-Goldstone (pNG) pion, leads to the vanishing mass in the chiral limit, but other 15 combinations remain heavy.
Although the pNG pion is customary treated as the pion, heavier pions do contribute to loop diagrams and thus induce lattice artifacts.
To reduce the lattice artifacts, various improved actions, including the asqtad, p4, stout and HISQ actions, have been proposed.

Because loop diagrams are the leading contributions in many thermodynamic quantities, a good control of the taste violation is important in finite temperature QCD.
In the studies of $T_c$, the main difference among the staggered-type quark actions seems to be the magnitude of the remaining taste violation: 
When we adjust the quark mass parameter to have $m_\pi^{\rm pNG} \approx 135$ MeV at lattice spacings corresponding to $T\approx170$ MeV on $N_t=8$ lattices, the masses for the heavy pions are around 400-600 (asqtad), 300-500 (stout) and 200-400 MeV (HISQ), respectively. 
The p4 is slightly worse than the asqtad.
Therefore, 
the p4 and asqtad suffers from severer lattice artifacts than the stout and HISQ.

The overall effects of heavy pions in the loop diagrams may be measured by a root mean squared (RMS) mass of pions $m_\pi^{\rm RMS}$.
In the right panel of Fig.~\ref{Fig:Tc1}, 
the asqtad data is shown to be well reproduced by the stout quark action by adjusting $m_\pi^{\rm RMS}$ \cite{WB_Tc3}.
This explains the larger $T_c$ with the p4 and asqtad actions at $N_t$ \lsim\ 8.

While the major part of the discrepancy is explained by the taste violation, the sensitivity of the determination of $T_c$ on the operator, observed with the stout quark, is not confirmed by other quarks yet.

In all of these studies, the physical point is identified by the pNG pion mass.
The success of the RMS pion mass suggests that it is more appropriate to consult the RMS hadron masses to judge the location of the physical point for thermodynamic quantities.
This will make the physical point closer to the chiral limit and the values of $T_c$ even smaller on finite lattices. 

\paragraph{Other lattice quarks}

It is highly desirable to confirm the results with other types of lattice quarks whose theoretical basis is solid.
With Wilson-type lattice quarks, the correct continuum limit is guaranteed for any number of flavors, 
at the price of explicit violation of the chiral symmetry at finite lattice spacings and more computational costs.
Due to the higher computational costs, the status of the studies with Wilson-type quarks is quantitatively much behind at present.
Recent updates of $T_c$ in two-flavor QCD are given in Refs.~\cite{ZeidlewiczLat10,Bornyakov09102392,WHOT_PRD82}.

The most attractive but the most expensive way is to adopt a chiral lattice quark action, such as the domain-wall and overlap.
The first estimate of $T_c$ with 2+1 flavors of domain-wall quarks is made by the HotQCD Collaboration \cite{ChengPRD81}.
The JLQCD collaboration has started finite temperature studies of QCD using an overlap quark action at a fixed topology \cite{CossuLat10}.


\section{Phase structure}

\begin{figure}
\hspace*{-4mm}
  \includegraphics[width=0.5\textwidth]{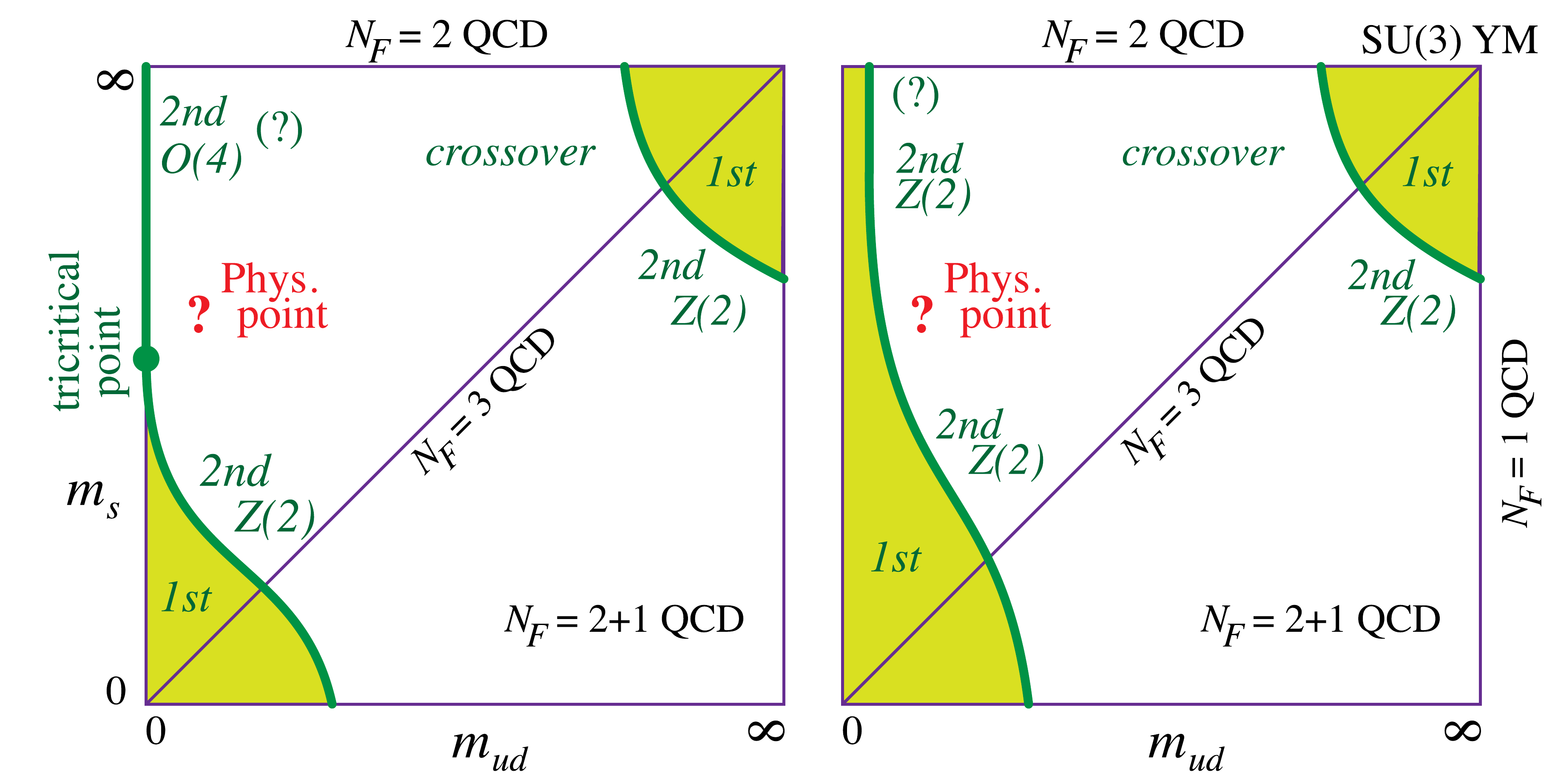}
\caption{Order of the finite temperature transition in 2+1 flavor QCD as a function of the degenerate $u$ and $d$ quark mass $m_{ud}$ and the $s$ quark mass $m_s$.
{\bf (Left)} 
The standard scenario with the second order chiral transition for two-flavor QCD. 
{\bf (Right)} 
An alternative scenario when the two-flavor chiral transition is first order.}
\label{Fig:Nf21PhaseDiagram}
\end{figure}

Figure~\ref{Fig:Nf21PhaseDiagram} summarizes the current wisdom about the order of the finite temperature transition in 2+1 flavor QCD as a function of the quark masses,
based on the studies on the lattice and with effective models.
When the $u$, $d$ and $s$ quarks are all sufficiently light or sufficiently heavy, we expect the transition to be of first order.
At intermediate values of quark masses, the ``transition'' will be an analytic crossover.
On the boundaries of the first order regions, we expect second order transitions.
Associated critical scaling around there will have characteristics universal to the corresponding effective models.

The nature of the transition in the chiral limit of two-flavor QCD (the upper left edge of the figure) has significant implications for the nature of the transition at the physical point too.
The transition is predicted to be either second order in the universality class of the O(4) Heisenberg model, or first order if the anomaly is weak around the transition point \cite{PisarskyWilczek}.

The left panel of Fig.~\ref{Fig:Nf21PhaseDiagram} summarizes the standard scenario in which the chiral transition of two-flavor QCD is second order.
In this case, because the chiral transition of three-flavor QCD is of first order, we have a tricritical point on the left edge of the figure ($m_{ud}=0$) where the order of the transition changes from the second order to the first order. 
Depending on the location of the tricritical point relative to the physical point, the universality class dominating the parameter dependence around the physical point will be different.

The right panel of Fig.~\ref{Fig:Nf21PhaseDiagram} shows an alternative scenario in which the chiral transition of two-flavor QCD is first order.
In this case, we have no tricritical point and thus no regions for the O(4) universality class.
A distinction between the two scenarios is important for studies at finite densities too.
Although the majority view the standard scenario as more probable, 
the nature of the two-flavor chiral transition was not fully fixed.
Here, scaling is a powerful tool to discriminate the nature of the transition.

\paragraph{O(4) scaling with Wilson-type quarks}

\begin{figure}[tbh]
\hspace*{-3mm}
  \includegraphics[width=0.245\textwidth]{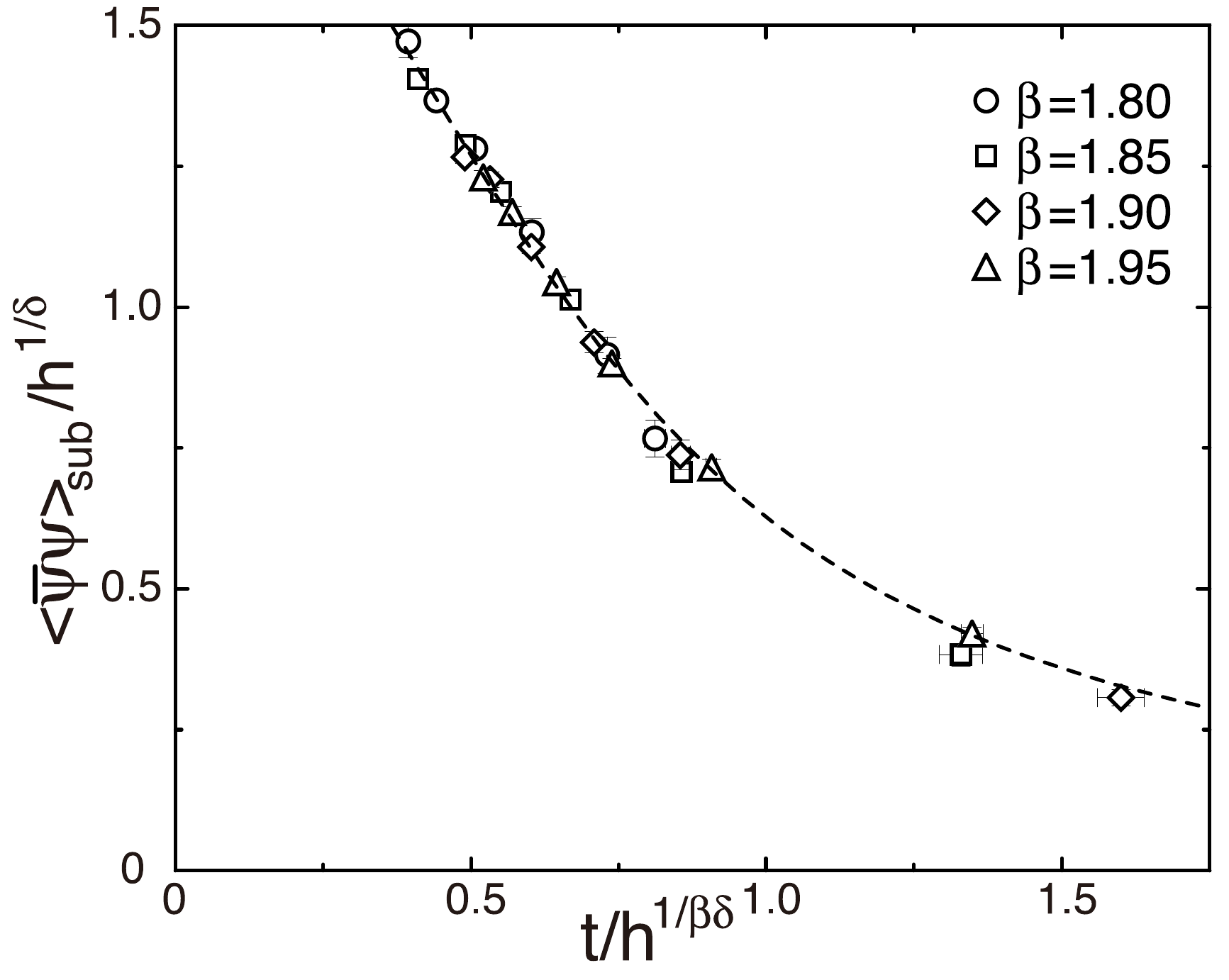}
  \includegraphics[width=0.24\textwidth]{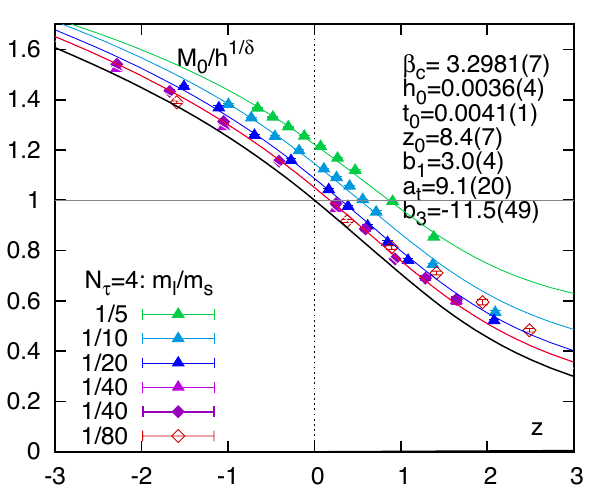}
\caption{
{\bf (Left)} 
O(4) scaling test for the subtracted chiral condensate in two-flavor QCD using the clover-improved Wilson quark action and the RG-improved Iwasaki gauge action on an $N_t=4$ lattice by the CP-PACS Collaboration\cite{CppacsPRD63}.
The dashed curve is the scaling function obtained in an O(4) spin model.
Similar result supporting the O(4) scaling was obtained with the unimproved Wilson quark action and the Iwasaki gauge action too \cite{IwasakiPRL78}.
{\bf (Right)} 
Rescaled chiral condensate $M_0 = m_s \langle \bar\psi\psi \rangle_{ud}/ T^4$ 
in 2+1 flavor QCD using the p4-improved staggered quark action and tree-level Symanzik gauge action 
on an $N_t=4$ lattice by the BNL-Bielefeld  Collaboration \cite{BNLB_O4}.
The QCD data are fitted to the O(2) scaling function with correction terms modeling a deviation from the scaling.
}
\label{Fig:O4W}
\end{figure}

With Wilson-type quarks, the O(4) scaling has been observed  for various actions since the early years of the investigations. 
Although Wilson-type quarks explicitly violate the chiral symmetry at finite lattice spacings, when the theory is correctly renormalized such that the chiral violations are subtracted out and the chiral Ward identities are satisfied, 
the resulting subtracted observables should show the correct chiral behaviors near the continuum limit \cite{Bochicchio}.

Adopting a chiral condensate defined by the axial vector Ward identity \cite{IwasakiPRL78},
the QCD data are shown to satisfy the O(4) scaling, with both unimproved and clover-improved Wilson quark actions \cite{IwasakiPRL78,CppacsPRD63,EjiriLat10}.
See the left panel of Fig.~\ref{Fig:O4W}.
A recent study on finer lattices with lighter quark masses supports the O(4) behavior too  \cite{Bornyakov09102392}.
These results, together with inconsistencies with first order scaling ans\"atze, suggest that the transition is continuous in the chiral limit.
However, the quark masses studied are heavy yet ($m_\pi$ \gsim\ 400 MeV).
Confirmation with lighter quarks on a fine lattice is indispensable.
Studies towards this direction are in progress (see e.g.\ \cite{MainzFrankfurtLat10}).

\paragraph{Chiral scaling with staggered-type quarks}

Scaling studies with staggered-type quarks have additional caveats.
With the fourth root procedure, the symmetry of the system at finite lattice spacings is identical to that of the original four-taste theory for any $N_F$.
In the chiral limit, we thus have the O(2) symmetry.
This means that, when the chiral transition is of second order, we should expect scaling properties in the universality class of O(2).%
\footnote{
In the studies of the O(2) scaling with staggered-type quarks, it is assumed that the non-locality of rooted staggered quark actions does not invalidate the universality arguments at finite lattice spacings too, 
though no theoretical argument supporting this assumption is known.
}
The O(4) scaling may appear only when we take the continuum limit prior to the chiral extrapolation.
We may, however, expect that when the chiral transition is continuous on finite lattices, it will remain so in the continuum limit.
In this sense, examination of O(2) is important, though the O(2) scaling is numerically not easy to discriminate from  the O(4) scaling.

The results for the chiral transition with two flavors of staggered quarks have been quite confusing.
Although the early studies suggest a continuous chiral transition, the scaling observed was neither O(2) nor O(4) \cite{Bielefeld94,Milc96,Jlqcd98}.
In contradiction to these studies, the Genova-Pisa group reported indications of a first order transition on an $N_t=4$ lattice \cite{Pisa}.

All of these studies are made using the unimproved staggered quark action and the unimproved plaquette gauge action, however.
As discussed in the previous section, large taste-violation with these actions will make the quark masses effectively much heavier than those naively estimated by the pNG pion mass.

This year, a new study with an improved staggered quark action was presented:
Using p4 quarks, the BNL-Bielefeld Collaboration performed a scaling test in 2+1 flavor QCD on an $N_t=4$ lattice 
for the quark mass range corresponding to $m_\pi^{\rm pNG} \approx 150$ MeV down to 75 MeV \cite{BNLB_O4}, which is much lighter than the previous studies.
For the first time, the O(2) scaling for staggered-type quarks has been observed, as shown in the right panel of Fig.~\ref{Fig:O4W}.
At the same time, they found substantial deviation from the O(2) scaling for $m_{ud}/m_s > 0.05$.
Together with the adoption of improved actions, this deviation explains the failures in previous studies with much heavier quark masses.
A preliminary result obtained on a finer $N_t=8$ lattice was shown to be similar \cite{SchmidtLat10}.

A big impact of this study is the O(2) dominance around the physical point.
First, this suggests that the chiral transition in two-flavor QCD is second order.
Second, the $s$ quark mass for the tricritical point may be smaller than the physical value.
In Fig.~\ref{Fig:Nf21PhaseDiagram}, I have therefore put the physical point slightly higher than the previous plots.
When the parameter-dependence around the physical point is confirmed to be dominated by the O(4) scaling in the continuum limit, the studies of QGP will be largely accelerated.

\paragraph{Near the heavy quark limit}

The upper right corner of Fig.~\ref{Fig:Nf21PhaseDiagram} corresponds to the heavy quark limit where the deconfinement transition of SU(3) pure gauge theory is described by the first order transition of an effective Z(3) Potts model \cite{YaffePRD26}. 
Dynamical heavy quarks affect as an external magnetic field to the effective spins and break the Z(3) symmetry explicitly.

The WHOT-QCD Collaboration studied the fate of the deconfinement transition when the quark mass is decreased from infinity \cite{SaitoLat10}.
Adopting a method using probability distribution of observables combined with a reweighting technique, they succeeded in calculating the effective potential, defined as the logarithm of the probability distribution for the plaquette, for a wide range of the plaquette expectation value. 
Close to the heavy quark mass limit,
the slope of the effective potential shows the shape like the letter "S" that is typical for a first order transition.
The ``S''-shape is shown to become weaker and eventually invisible when $m_q$ is decreased.
Determining the location of the critical point terminating the first order transition, they obtained the phase diagram for 2+1 flavor QCD.
As naively expected, the endpoint monotonically shifts to lighter quark mass as we increase the number of flavors.


\section{Equation of state}

Calculation of the EOS in full QCD is demanding.
Besides the basic information such as the lattice scale, the non-perturbative beta functions and  the line of constant physics (LCP), we need the expectation values at $T=0$ to renormalize finite temperature observables at each simulation point.
Conventionally, we vary the coupling parameters along a LCP with a fixed temporal lattice size $N_t$ to vary the temperature (fixed $N_t$ approach).
To calculate the EOS non-perturbatively, the integration method is adopted \cite{Integral}.
These require a systematic scan in a wide range of parameters on both zero and finite temperature lattices
with sufficiently large spatial sizes.
For a continuum extrapolation, we repeat the calculation at several values of $N_t$. 

\paragraph{EOS with 2+1 flavors of improved staggered quarks}

The BNL-Bielefeld Collaboration published the EOS ``with physical quark masses'', based on their simulations using  p4 quarks on an $N_t=8$ lattice \cite{BNLB_EOSp}.
Using a scale set by $r_0$, their simulation point at the bare quark mass ratio $m_{ud}/m_s = 0.05$ corresponds to $m_\pi^{\rm pNG} =154$ MeV.
The Wuppertal-Budapest Collaboration calculated the EOS with stout quarks at the physical point on $N_t=6$--10 lattices (and a few points at $N_t=12$) \cite{WB_EOS10}.
They also attempted to calculate the EOS up to $T =1000$ MeV
using LCP determined by a step-scaling method.
At Lattice 2010, the HotQCD Collaboration presented preliminary results of  the EOS with HISQ and asqtad quarks at $m_{ud}/m_s = 0.05$ on $N_t=8$ lattices \cite{BazavovSoeldnerLat10}.

\begin{figure}[tbh]
\hspace*{-2mm}
  \includegraphics[width=0.245\textwidth]{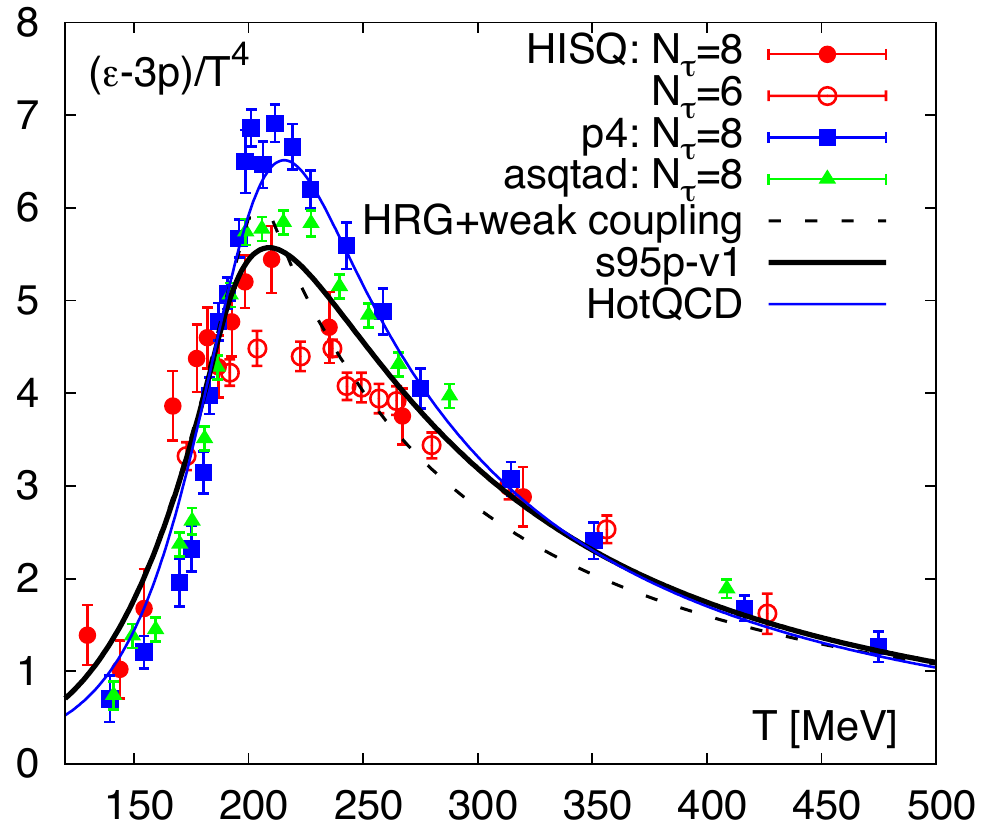}
\hspace*{-2mm}
  \includegraphics[width=0.245\textwidth]{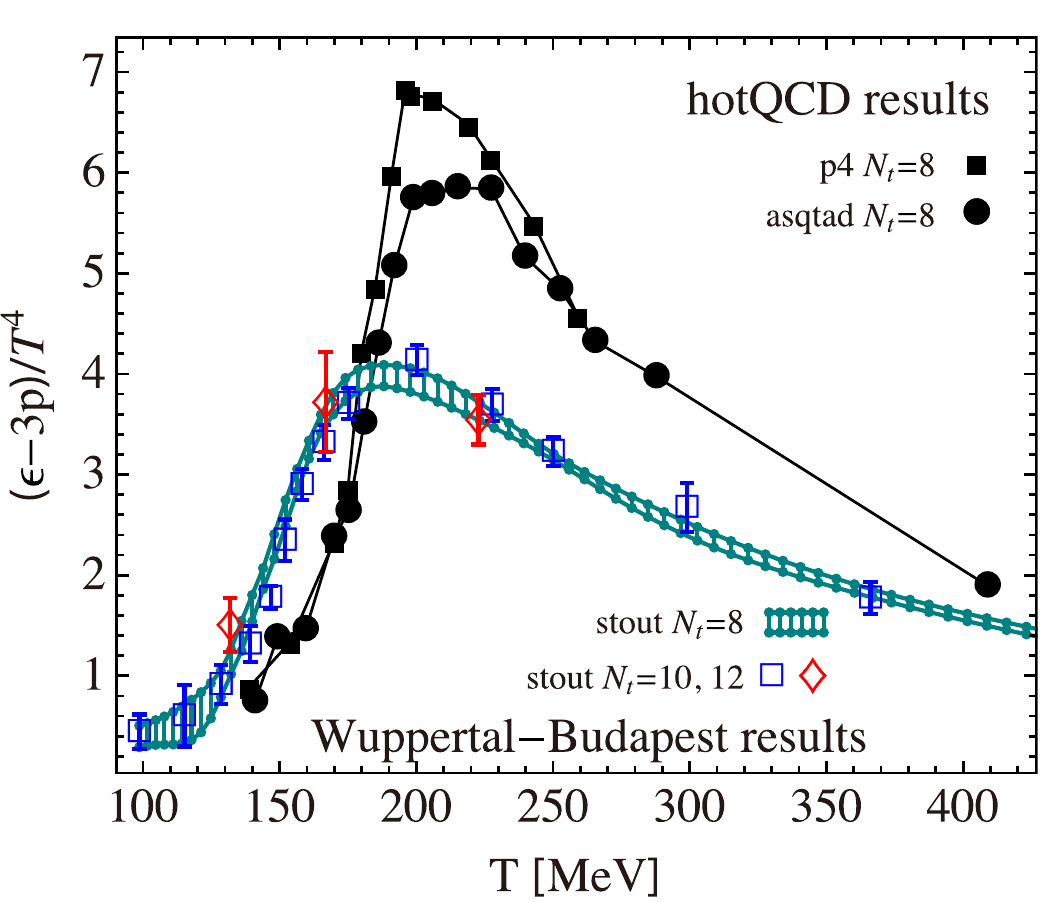}
\caption{
{Trace anomaly in 2+1 flavor QCD.
The scale is set by $r_0$, except for the stout data for which the scale is set by $f_K$.
\bf (Left)}
Comparison of HSQ, asqtad and p4 quarks at $m_{ud}/m_s = 0.05$ on $N_t=8$ lattices \cite{BazavovSoeldnerLat10}.
{\bf (Right)}
Comparison of stout, asqtad and p4 quarks on lattices $N_t \ge 8$ \cite{WB_EOS10}.
The stout data is obtained at the physical point ($m_{ud}/m_s \approx 0.035$) while asqtad and p4 data are for $m_{ud}/m_s = 0.05$.
The stout values are corrected by a tree-level improvement factor.
}
\label{Fig:EOSS1}
\end{figure}

\begin{figure}[tbh]
\hspace*{-2mm}
  \includegraphics[width=0.195\textwidth]{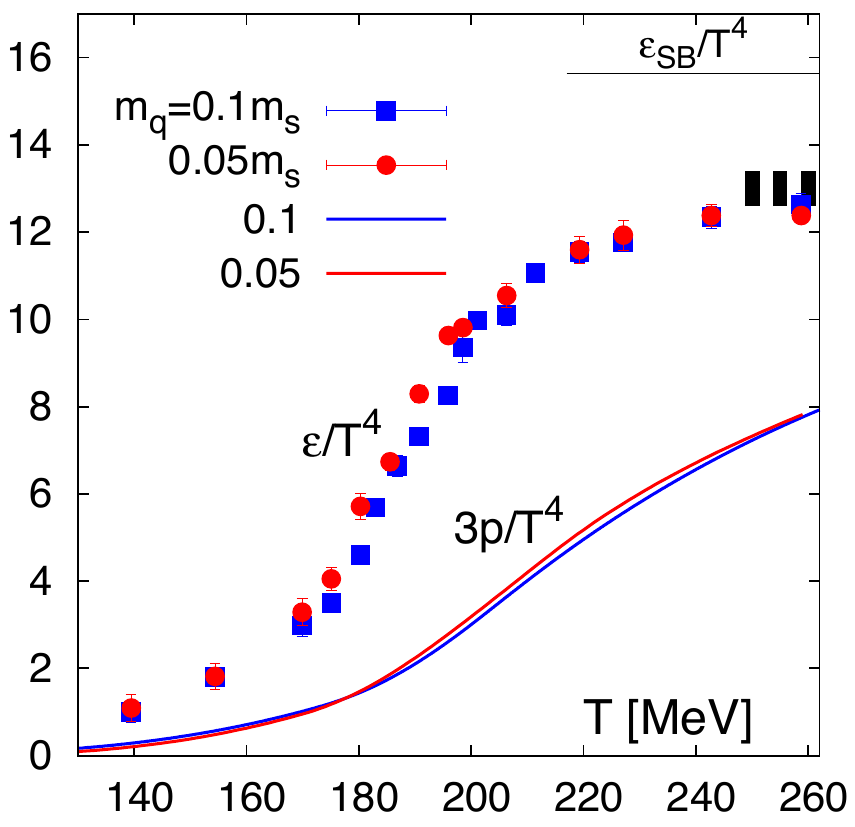}
\hspace*{-2mm}
  \includegraphics[width=0.285\textwidth]{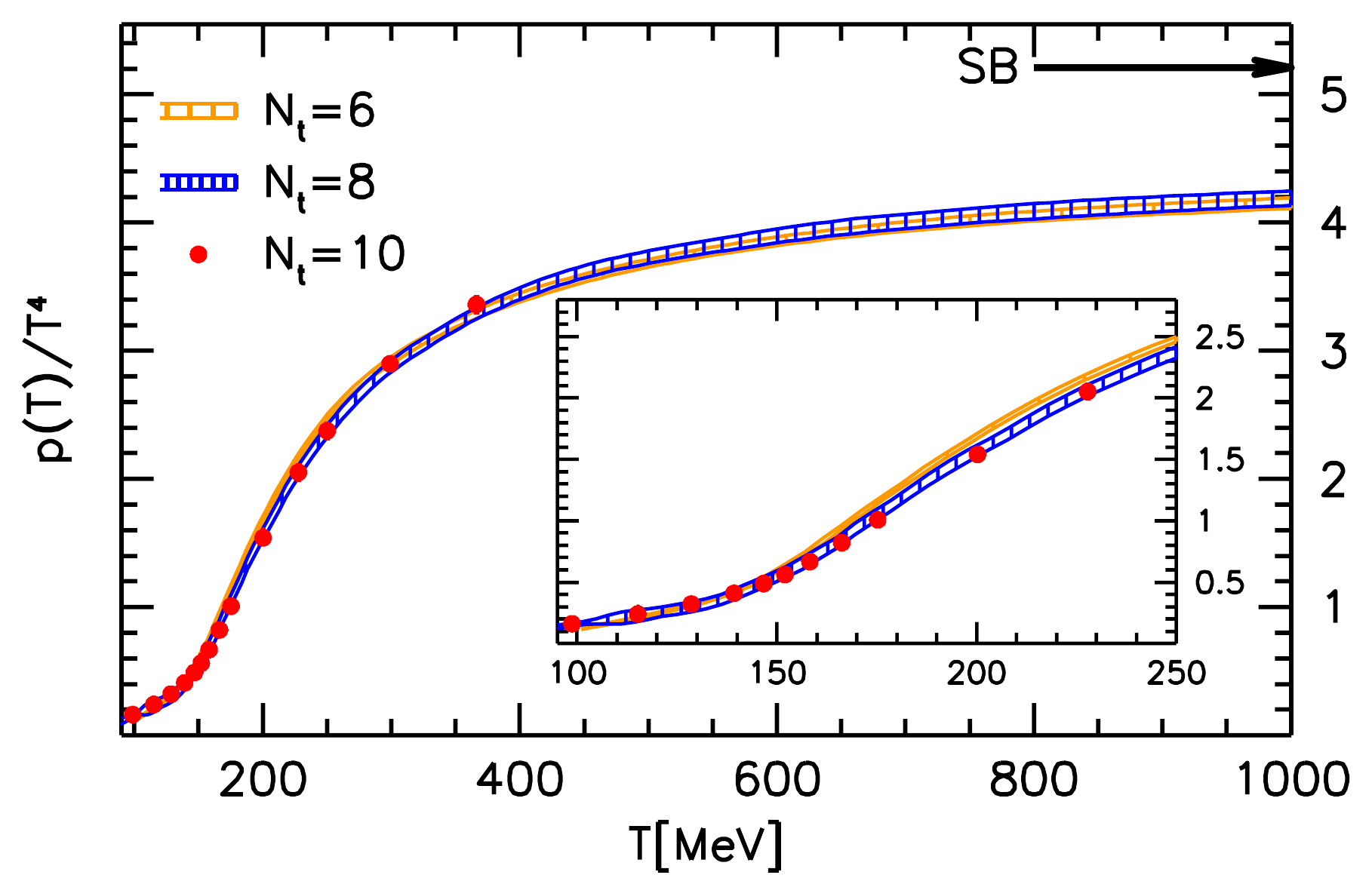}
\caption{EOS in 2+1 flavor QCD.
{\bf (Left)}
p4 results at $m_\pi^{\rm pNG} \approx 154$ MeV ($m_{ud}/m_s = 0.05$) obtained on an $N_t=8$ lattice using the $r_0$ scale \cite{BNLB_EOSp}.
Results at $m_\pi^{\rm pNG} = 220$ MeV ($m_{ud}/m_s = 0.1$) are compared.
{\bf (Right)}
Pressure with stout quarks using the $f_K$ scale \cite{WB_EOS10}.
}
\label{Fig:EOSS2}
\end{figure}

The results for the trace anomaly from these studies are plotted in Fig.~\ref{Fig:EOSS1}. 
In the left panel, a comparison is made at  $m_{ud}/m_s = 0.05$ for HSQ, asqtad and p4 quarks.
In the right panel, the stout data is compared with p4 and asqtad results.
While the location of the transition is approximately consistent with each other, 
the magnitude of the trace anomaly shows clear dependence on the action adopted.
In particular, we note that the peak of the trace anomaly around $T \sim 200$ MeV becomes lower with the adoption of a more improved action.

Using the integration method, the EOS around the physical point is obtained, as shown in Fig.~\ref{Fig:EOSS2}.
Note that, in these studies, the physical point was identified by the pNG pion mass, and the chiral and continuum extrapolations are not done yet.

\paragraph{EOS with 2+1 flavors of improved Wilson quarks}

About ten years ago, the CP-PACS Collaboration performed a systematic calculation of the EOS in two-flavor QCD with clover-improved Wilson quarks coupled to RG-improed Iwasaki glue \cite{CppacsPRD63,CppacsPRD64}.
Extension of the study to 2+1 flavor QCD was not straightforward due to the heavy computational demand for Wilson-type quarks.

To overcome the problem, the WHOT-QCD Collaboration developed a new approach: a fixed scale approach \cite{Tintegral}
in which $T$ is varied through a variation of $N_t$ with the coupling parameters fixed.
The conventional integration method \cite{Integral} is inapplicable in the fixed-scale approach. 
Therefore, they developed an alternative integration method, the $T$-integration method \cite{Tintegral}. 
With the fixed scale approach, the cost for zero-temperature simulations can be largely reduced, because a common zero-temperature result can be used to the renormalization at all temperatures.
We may even borrow high statistics zero-temperature configurations publicly available on the International Lattice Data Grid (ILDG).
The fixed-scale approach is complementary to the fixed $N_t$ approach.
Towards the high $T$ limit, the fixed-scale approach suffers from lattice artifacts due to small $N_t$, while the fixed $N_t$ approach is applicable for quantities insensitive to the system volume.
In the low $T$ region, the fixed $N_t$ approach suffers from coarseness of the lattice, while the fixed-scale approach keeps the lattice spacing.

\begin{figure}[tbh]
  \includegraphics[width=0.295\textwidth]{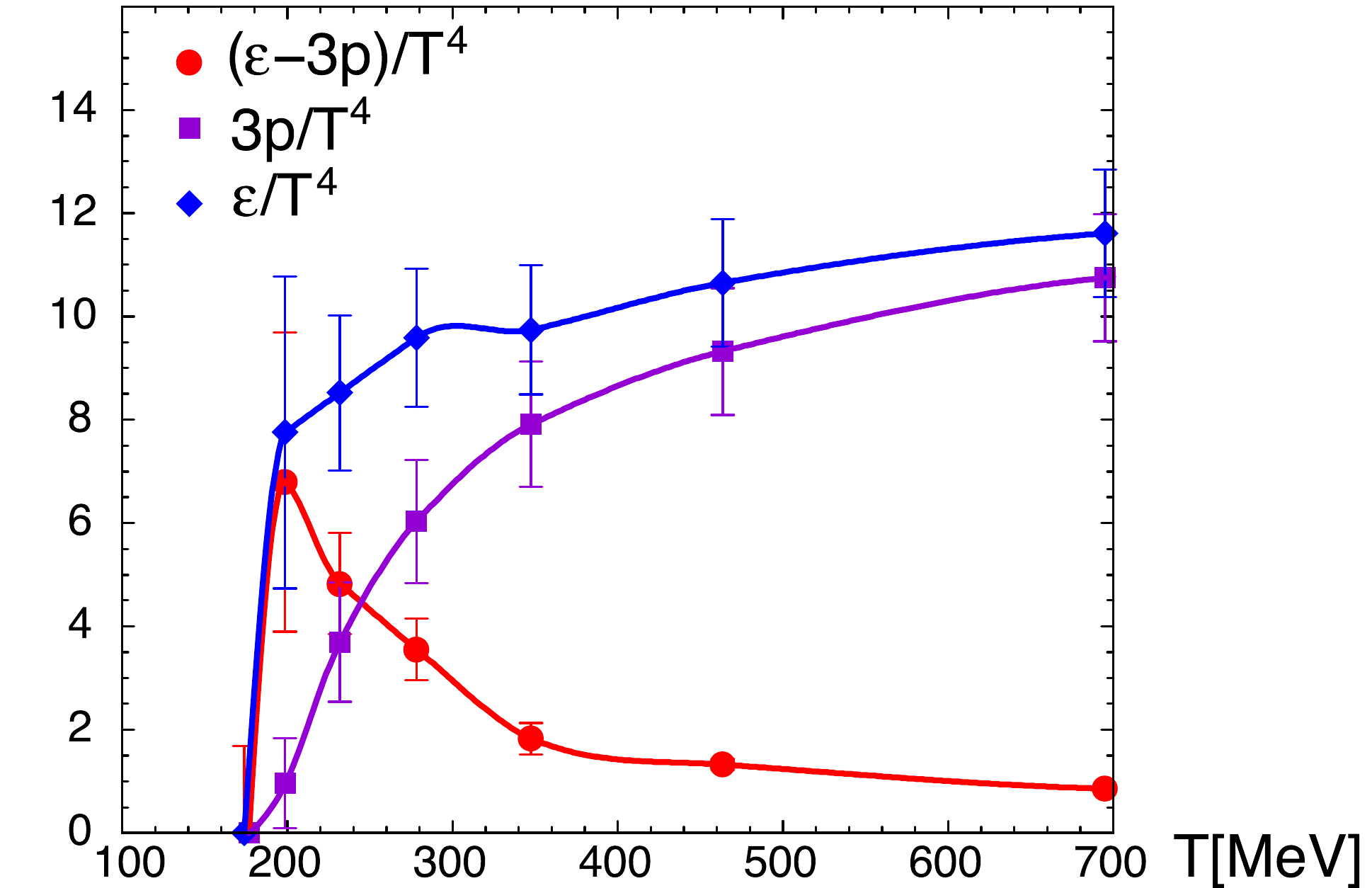}
\caption{
EOS in 2+1 flavor QCD with clover-improved Wilson quarks adopting the fixed-scale approach and the $T$-integration method \cite{UmedaLat10}.
}
\label{Fig:EOSW}
\end{figure}

Adopting the fixed-scale approach, the WHOT-QCD Collaboration attempted the first calculation of the EOS in 2+1 flavor QCD with Wilson-type quarks \cite{UmedaLat10}.
They borrowed zero-temperature configurations using the non-perturbatively improved clover quark action \cite{CppacsJlqcdPRD78}, 
and performed finite temperature simulations on $N_t=4$--16 lattices, which corresponds to $T\approx 170$--700 MeV.
Their preliminary result for the EOS is shown in Fig.~\ref{Fig:EOSW}.
The peak height of the trace anomaly, obtained on $N_t \approx 14$ lattices, is roughly consistent with those shown in Fig.~\ref{Fig:EOSS1} with highly improved staggered quarks, though the statistical error is large.
Further statistics are needed  to reduce the errors.


\section{Summary}

There have been important advances in finite temperature QCD on the lattice in the past years.
Large-scale simulations of 2+1 flavor QCD with various improved staggered quark actions clarified the origin of the recent discrepancy in the transition temperature.
From this, we have learned the importance of the effect of taste violation in the studies with staggered-type quarks.
The O(2) chiral scaling for staggered-type quarks was observed by reducing the light quark mass down to a quite low value.
At the same time, the theoretical uneasiness with staggered-type lattice quarks motivated several groups to accelerate studies adopting Wilson-type and lattice chiral quark actions. 
Calculation of the EOS in 2+1 flavor QCD has started with improved Wilson quarks.
Furthermore, simulations with domain-wall quarks have entered a level of quantitative investigations.


\begin{theacknowledgments}
I thank the organizers of Confinement IX for stimulating conference.
I am grateful to the colleagues in the lattice society for sending me results and comments. 
This work is in part supported by Grants-in-Aid of the Japanese Ministry of Education, Culture, Sports, Science and Technology (No. 21340049).
\end{theacknowledgments}


\end{document}